\documentclass{article}

\usepackage{arxiv}

\usepackage[english]{babel}
\usepackage[utf8]{inputenc} 
\usepackage[T1]{fontenc}    
\usepackage{hyperref}       
\usepackage{url}            
\usepackage{booktabs}       
\usepackage{amsfonts}       
\usepackage{nicefrac}       
\usepackage{microtype}      
\usepackage{lipsum}

\usepackage{xcolor}
\usepackage{graphicx}
\usepackage{multirow}
\usepackage{tabularx}
\usepackage{amsmath}
\usepackage[ruled,vlined]{algorithm2e}

\author{
  \textbf{Dimitri Leandro de Oliveira Silva} \\
  Laboratory of Signals and Systems \\
  Federal University of ABC\\
  Santo André, SP - Brazil \\
  \texttt{dimitri.leandro@aluno.ufabc.edu.br} \\
  \and
  \textbf{Tito Spadini} \\
  Laboratory of Signals and Systems \\
  Federal University of ABC\\
  Santo André, SP - Brazil \\
  \texttt{tito.caco@ufabc.edu.br} \\
  \and
  \\ \textbf{Ricardo Suyama} \\
  Laboratory of Signals and Systems \\
  Federal University of ABC\\
  Santo André, SP - Brazil \\
  \texttt{ricardo.suyama@ufabc.edu.br}
}

\title{Microphone Array Based Surveillance Audio Classification}

\begin{document}
\maketitle
\begin{abstract}
The work assessed seven classical classifiers and two beamforming algorithms for detecting surveillance sound events. The tests included the use of AWGN with -10 dB to 30 dB SNR. Data Augmentation was also employed to improve algorithms' performance. The results showed that the combination of SVM and Delay-and-Sum (DaS) scored the best accuracy (up to 86.0\%), but had high computational cost ($\approx $ 402 ms), mainly due to DaS. The use of SGD also seems to be a good alternative since it has achieved good accuracy either (up to 85.3\%), but with quicker processing time ($\approx$ 165 ms).
\end{abstract}


\section{Introduction}
    \label{sec:introducao}

    Several public security systems depend directly on human action in numerous stages of its operation. The monitoring of public areas, for instance, is usually done with the use of cameras spread over the busiest places in large urban centers. In general, these systems depend on an operator to pay attention to the images so that the agencies responsible for security can be activated when events such as thefts, vandalism, and traffic accidents are observed. Considering the amount of information to which the operator is exposed, there is a high probability that surveillance failures will occur, even if the patrol center has a large team \cite{CROCCO2016}. Although the operators are attentive at all times, this type of monitoring has some disadvantages: the images are limited to the direction in which the camera points and have low visibility at dusk and in cases of rain or bright light. Besides, events such as gunshots, alarms, distress calls, among others, are much more noticeable in the auditory field than in the visual \cite{RATY2010, 11-MULIMANI2019}.
    
    In this sense, the monitoring of risk areas could be done through the use of audio processing techniques, reducing the need for human participation in the surveillance process, and making public security systems more efficient \cite{17-HASSAN2019}. To support this argument, it is worth recalling two very favorable characteristics concerning these signals: initially, the sound consumes less bandwidth in the transmission of information, reducing the need for high transmission rates, as in the case of high definition images; in addition, sound processing techniques require, in general, less computational power than techniques for video processing and analysis, which would enable the implementation of simpler and therefore less costly embedded systems \cite{11-MULIMANI2019, 20-BELLOCH2020}.
    
    \begin{figure}[!htb]
        \centering
        \includegraphics[keepaspectratio, width=10cm]{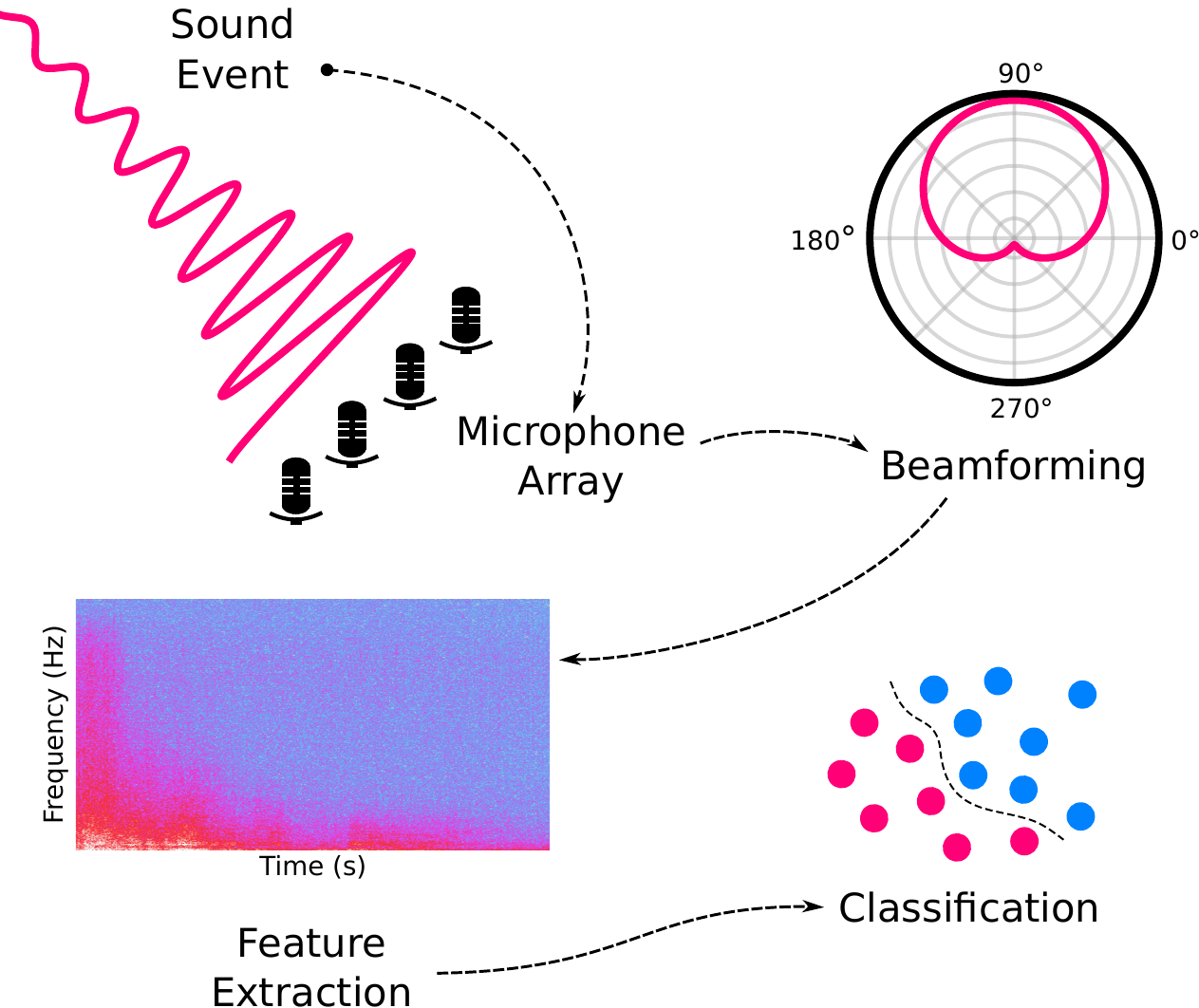}
        \caption{Illustration of the steps of the proposed system: from audio capture to classification.}
        \label{fig:propostaSistema}
    \end{figure}
    
    The present work is a sequence of what is presented in \cite{spadini2019sound}, and aims to evaluate the effectiveness of audio processing techniques for detecting events that are harmful to public safety, based on two important points: the use of machine learning techniques for automatic recognition of sound events; and signal processing in microphone arrays, to improve the signal-to-noise ratio (SNR) and, therefore, obtain better accuracy in the classification stage. Figure \ref{fig:propostaSistema} summarizes the proposed system.
    
    Following, Section \ref{sec:revisao} presents a bibliographic review on the subjects covered in this article, and Section \ref{sec:algoritmos} reveals the main considerations about the algorithms employed. Section \ref{sec:ambiente} explains the simulation environment and procedures adopted, while Section \ref{sec:resultados} discuss the outcomes. Finally, Section \ref{sec:conclusao} concludes the work.

\section{Literature Review}
    \label{sec:revisao}
    
    In recent years, researchers have shown that the most effective tools for the classification of sound events include the application of deep, convolutional, and recurrent neural networks (DNN, CNN, and RNN) \cite{7-VECCHIOTTI2019, 9-SU2020, 11-MULIMANI2019, 15-BAI2018, 17-HASSAN2019}. However, for the current work, the concern with the processing time of the algorithms is fundamental, since, among the future goals, the aim is to create a low-cost system capable of running in real-time. Therefore, the solutions used here address classic machine learning algorithms, seeking to find a balance between accuracy and computational cost. In \cite{16-SURAMPUDI2019}, the authors also worked with audio classification for security systems and used two of the seven algorithms that will be covered in this article, K Nearest Neighbors (KNN) and Support Vector Machines (SVM). Similar to the previous one, \cite{11-MULIMANI2019} revealed that Random Forests can achieve high accuracy with MIVIA Dataset \cite{miviaDataset}, specifically for surveillance applications, even with low SNR.
    
    Recent work advocates the use of DNN and CNN can perceive patterns in auditors without using many features \cite {15-BAI2018, 7-VECCHIOTTI2019}. Both were able to acquire good results using only Mel Frequency Cepstral Coefficients (MFCC). \cite{9-SU2020, 11-MULIMANI2019, 17-HASSAN2019} states that temporal and frequency features, when separately used, can't achieve satisfactory performance, especially in noisy environments, but the combination of those can significantly improve the classification task. The authors \cite{9-SU2020} and \cite{17-HASSAN2019} claim that Spectral Chroma, Spectral Contrast, and Tonnetz were primarily related to musical classification, but examine their performance in classification of other sorts of audio signals, attesting they exert a fundamental role in that duty. In \cite{9-SU2020, 11-MULIMANI2019}, and \cite{17-HASSAN2019}, the authors confirm the excellent performance of MFCC and its first and second-order derivatives, Delta and Delta Delta.

    In \cite{1-SCARDAPANE2015} and \cite{15-BAI2018}, the authors use beamforming to improve the classification of audios. The first one operates in the field of monitoring systems and, as in the current work, simulates an array of microphones and establishes the position of the sound source randomly. The second one seeks to classify the pronunciation of ten simple English words with 10~dB SNR and fixed position. As in \cite{15-BAI2018}, \cite{3-MUROVEC2018} argues that the use of beamforming improves the performance of classifiers, but as long as the direction of the emitting source is known. More recently, authors have used beamforming to ascertain the influence of different sound sources on the total noise of an environment surrounded by factories, highways and train stations, making it possible to determine the most intense source \cite{3-MUROVEC2018}. In all articles mentioned, the authors apply the Delay-and-Sum (DaS) algorithm and achieve good classification performance. In \cite{20-BELLOCH2020}, a real-time system for locating sound sources in the ESP32 micro-controller was developed, exposing the feasibility of future works mentioned above.
    
    The current work leads contributions by addressing several procedures separately used in other articles, attempting to unify techniques and results that can be applied to a common objective. The report discusses comparatively the use of beamforming and machine learning techniques, relying on simulations of diffuse noise in a wide SNR range, seeking results both in accuracy and in processing time.

\section{Algorithms}
    \label{sec:algoritmos}
    
    This section will briefly present the main fundamentals of beamforming and machine learning algorithms employed in this work.
    
    \subsection{Beamforming}
    
    Beamforming algorithms are based on the fact that a wave propagating in a certain direction of space will be captured at different times by each device present in an array of receivers (microphones, in the case of this work) \cite{KRIM1996}. The simplest of them, Delay-and-Sum, does exactly what its name says: settle the delays between each microphone and then add them up. This causes directivity patterns to be created, amplifying signals from a certain direction, and attenuating from others \cite{mccowan2001, ZENG2014}. To find the lags of each microphone concerning the referential, auto-correlation analysis in the time or frequency domain can be performed \cite{KNAPP1976, 7-VECCHIOTTI2019}.

    \begin{figure}[!htb]
        \centering
        \includegraphics[keepaspectratio, width=10cm]{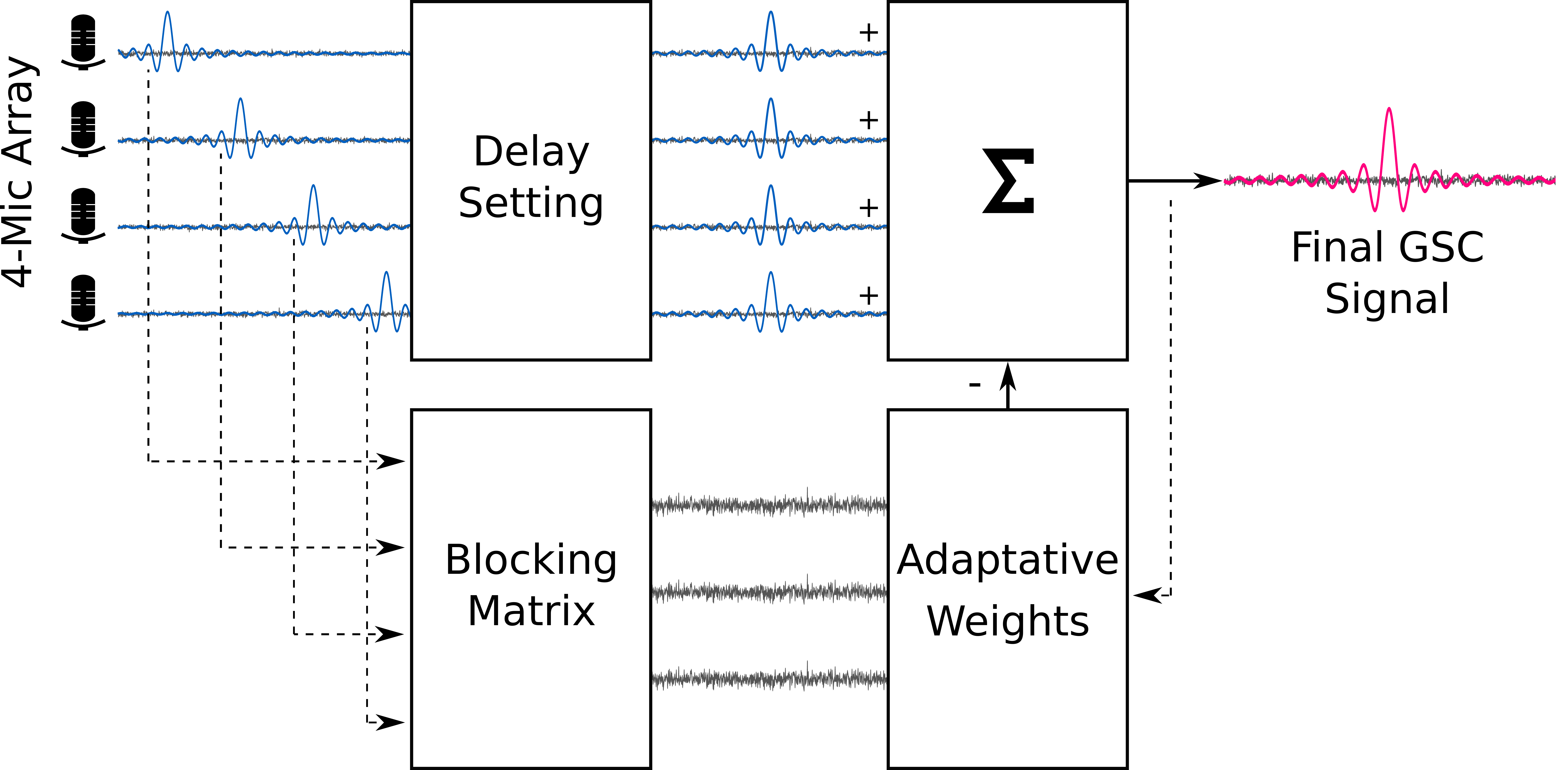}
        \caption{GSC structure adopted in this work, that considers an arrangement of four microphones.}
        \label{fig:esquemaGSC}
    \end{figure}
    
    The so-called Generalized Sidelobe Canceller differs from the previous one for being adaptive. It attempts to minimize the power of noise in the resulting signal coming from some other fixed beamforming algorithm, such as DaS itself. To achieve this goal, GSC drops the signal of interest from the signal captured by each microphone using a Blocking Matrix. Subsequently, these signals are subtracted from the final signal using an adaptive multiplicative factor \cite{mccowan2001, ROMBOUTS2008}. Figure \ref{fig:esquemaGSC} shows the structure of the GSC, and the Equation \ref{eq:blockingMatrix} shows the Blocking Matrix employed in this work. The algorithm used in the adaptive section was the Least Mean Squares (LMS).
    
    \begin{equation}
        B = 
        \begin{bmatrix}
        1 & -1 & 0  & 0 \\
        0 & 1  & -1 & 0 \\
        0 & 0  & 1  & -1 \\
        \end{bmatrix}
        \label{eq:blockingMatrix}
    \end{equation}

    \subsection{Classifiers}
    
    Supervised learning algorithms are based on prior knowledge of training samples and their respective labels. Therefore, given a new sample to be classified, the algorithm must stipulate its label according to the already known data, mathematically described by a vector of characteristics (features) \cite{Friedman2001}.
    
    In this work, seven classifiers will be addressed: K Nearest Neighbors (KNN), Linear Discriminant Analysis (LDA), Perceptron, Quadratic Discriminant Analysis (QDA), Stochastic Gradient Descent (SGD), Support Vector Machine (SVM) and Decision Tree \cite{Friedman2001}.

    \section{Simulation Environment and Procedures}
    \label{sec:ambiente}
    
     This section will describe the dataset and the procedures that were adopted to: increase the database in order to improve the performance of classifiers in noisy environments; find the best configuration for each classifier, so that these algorithms achieve superior performance, and; simulate a microphone array for each audio in the dataset with different SNR values. In all stages of this work, Python was employed in an Intel (R) Xeon (R) E5-2650 v3 @ 2.30~GHz CPU and 16~GB of RAM.
    
    \subsection{Dataset}
    
        The dataset Sound Events for Surveillance Applications (SESA) \cite{sesaDataset} is composed of 480 audios up to 33 seconds length, divided between the classes \textit{shot}, \textit{explosion}, \textit{alarm}, and \textit{casual}, the latter being composed of audios that could be incorrectly classified as any of the other three classes, but which do not represent security risks, e.g. constructions, fireworks, thunders, horns, among others. All audios are WAV files in mono, sampled at 16~kHz with 8 bits of depth.
        
    
        Attempting to improve the performance of the classifiers in scenarios in which the noise stands out from the signal of interest, the dataset underwent a process of Data Augmentation. The procedure consisted of adding white Gaussian noise to each audio in the dataset, with SNR ranging from -10 ~ dB to 30 ~ dB with step 5 ~ dB.
    
    \subsection{Features Extraction}
        
\begin{table}[!htb]
    \centering
    \caption{Features vector of each audio window.}
    
    \begin{tabular}{ll}
        \hline
            \textbf{Feature}        & \textbf{Position} \\
        \hline
            \rule{0pt}{2.0ex}
            Root Mean Square                 & [0] \\
            \rule{0pt}{2.0ex}
            Spectral Centroid            & [1] \\
            \rule{0pt}{2.0ex}
            Spectral Bandwidth              & [2] \\
            \rule{0pt}{2.0ex}
            Spectral Flatness                & [3] \\
            \rule{0pt}{2.0ex}
            Roll-Off Frequency              & [4] \\
            \rule{0pt}{2.0ex}
            Zero Crossing Rate                & [5] \\
            \rule{0pt}{2.0ex}
            MFCC           & [6--24] \\
            \rule{0pt}{2.0ex}
            Delta          & [25--44] \\
            \rule{0pt}{2.0ex}
            Delta Delta         & [45--64] \\
            \rule{0pt}{2.0ex}
            Mel Spectrogram      &      [65--74] \\
            \rule{0pt}{2.0ex}
            Chromagram         & [75--86] \\
            \rule{0pt}{2.0ex}
            Constant-Q Chromagram            & [87--98] \\
            \rule{0pt}{2.0ex}
            Chroma Energy Normalized (CENs)          & [99--110] \\
            \rule{0pt}{2.0ex}
            Tonnetz          & [111--117] \\
            \rule{0pt}{2.0ex}
            Spectral Contrast          & [118--125] \\
        \hline
    \end{tabular}

    \label{tab:vetorFeatures}
\end{table}
        
        In order to perform the features extraction procedure, the files were segmented into 200~ms and 50~\% overlap windows. Each window was designed as a feature vector, as shown in Table \ref{tab:vetorFeatures}, selected based on the discussions related to the works mentioned. After extraction, features were normalized. The work \cite{16-SURAMPUDI2019} explains the importance of this stage for the performance of the classifiers.
        
    \subsection{Gridsearch}
    
        Seven classifiers were submitted to Gridsearch, a process that performs, for each classifier, a search for the hyper-parameter values that give the best classification performance. The cross-validation method used was Bootstrap, which consists of several reproductions of separation of data between training and testing, to obtain more reliability in the results achieved with the classifiers \cite{Friedman2001}. The mode of each window classification determines the final classification of single audio. It is important to regard that Gridsearch was performed twice: first using the original dataset in the training phase; then, the augmented.

    \subsection{Microphone Arrangement Simulation}
    
        \begin{figure}[!htb]
            \centering
            \includegraphics[keepaspectratio, width=10cm]{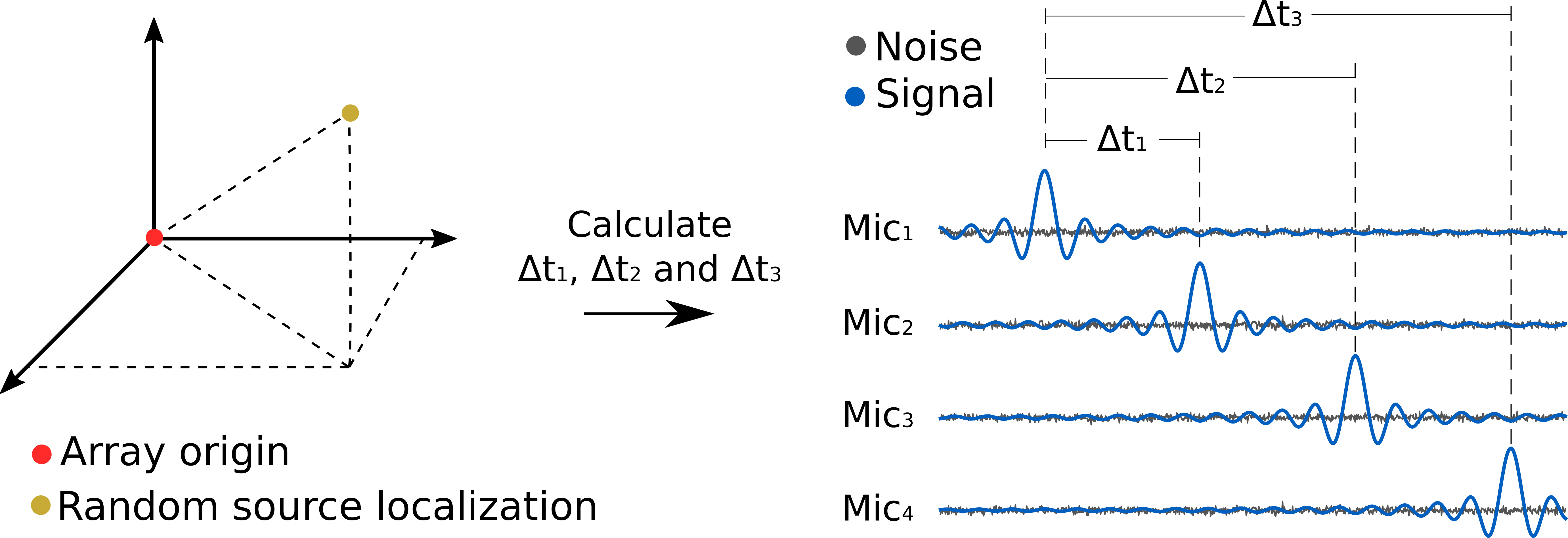}
            \caption{Procedure adopted to simulate an array of microphones with diffuse noise for each audio and for each SNR.}
            \label{fig:simulacaoArrayMics}
        \end{figure}
    
        After checking the best settings for the classifiers used in the original and augmented datasets, a procedure was used to simulate a microphone array so that these techniques could be evaluated. For this, the position of each microphone was designed to respect the positioning characteristics observed in the device called \textit{ReSpeaker 4-Mic Array}, as it will be used in continuations of this work \footnote{http://wiki.seeedstudio.com/ReSpeaker\_4\_Mic\_Array\_for\_Raspberry\_Pi/}.
    
        
        For each dataset audio, both azimuth and elevation angles were randomly defined so that the lags of each microphone concerning the reference could be calculated according to the array geometry. Following that, the audios were replicated with these same lags. Later on, white noise was added to each replicated signal, simulating a diffuse noise field \cite{mccowan2001}. This procedure was repeated for SNR values from -10~dB to 30~dB, with a step of 1~dB, as shown in Figure \ref{fig:simulacaoArrayMics}.
        
\section{Discussion and Analysis of the Results}
    \label{sec:resultados}
    
\begin{table*}[!ht]
    \centering
    \caption{Mean accuracy and standard deviation achieved before and after data augmentation.}
    
    \resizebox{\textwidth}{!}{
        \begin{tabular}{|l|l|c|c|c|c|c|c|c|c|c|c|}
\hline
\multicolumn{1}{|c|}{\multirow{2}{*}{\textbf{Classifier}}} & \multicolumn{1}{c|}{\multirow{2}{*}{\textbf{Beamforming}}} & \multicolumn{5}{c|}{\textbf{Original Dataset}}                                                                                  & \multicolumn{5}{c|}{\textbf{Data Augmentation}}                                                                                 \\ \cline{3-12} 
\multicolumn{1}{|c|}{}                                     & \multicolumn{1}{c|}{}                                     & \textit{\textbf{-10 dB}} & \textit{\textbf{0 dB}} & \textit{\textbf{10 dB}} & \textit{\textbf{20 dB}} & \textit{\textbf{30 dB}} & \textit{\textbf{-10 dB}} & \textit{\textbf{0 dB}} & \textit{\textbf{10 dB}} & \textit{\textbf{20 dB}} & \textit{\textbf{30 dB}} \\ \hline
\multirow{3}{*}{KNN}                                       & No Beamforming                                          & 0.605 $\pm$ 0.021            & 0.649 $\pm$ 0.033          & 0.710 $\pm$ 0.016           & 0.784 $\pm$ 0.017           & 0.828 $\pm$ 0.012           & 0.697 $\pm$ 0.018            & 0.796 $\pm$ 0.014          & 0.847 $\pm$ 0.019           & 0.840 $\pm$ 0.011           & 0.840 $\pm$ 0.007           \\ \cline{2-12} 
                                                           & DaS                                                 & 0.607 $\pm$ 0.020            & 0.691 $\pm$ 0.025          & 0.761 $\pm$ 0.012           & 0.851 $\pm$ 0.021           & \textbf{0.878 $\pm$ 0.007}  & 0.769 $\pm$ 0.016            & 0.820 $\pm$ 0.007          & 0.836 $\pm$ 0.009           & 0.840 $\pm$ 0.011           & 0.849 $\pm$ 0.003           \\ \cline{2-12} 
                                                           & GSC                                                       & 0.619 $\pm$ 0.013            & 0.695 $\pm$ 0.015          & 0.754 $\pm$ 0.018           & 0.855 $\pm$ 0.019           & 0.868 $\pm$ 0.013           & 0.750 $\pm$ 0.018            & 0.822 $\pm$ 0.004          & 0.838 $\pm$ 0.019           & 0.855 $\pm$ 0.015           & 0.838 $\pm$ 0.006           \\ \hline
\multirow{3}{*}{LDA}                                       & No Beamforming                                          & 0.542 $\pm$ 0.034            & 0.670 $\pm$ 0.018          & 0.721 $\pm$ 0.026           & 0.720 $\pm$ 0.017            & 0.702 $\pm$ 0.007           & 0.622 $\pm$ 0.011            & 0.697 $\pm$ 0.013          & 0.739 $\pm$ 0.011           & 0.786 $\pm$ 0.012           & 0.744 $\pm$ 0.007           \\ \cline{2-12} 
                                                           & DaS                                                 & 0.580 $\pm$ 0.015            & 0.681 $\pm$ 0.009          & 0.744 $\pm$ 0.021           & 0.788 $\pm$ 0.003           & 0.809 $\pm$ 0.017           & 0.645 $\pm$ 0.011            & 0.716 $\pm$ 0.018          & 0.731 $\pm$ 0.016           & 0.767 $\pm$ 0.009           & 0.777 $\pm$ 0.007           \\ \cline{2-12} 
                                                           & GSC                                                       & 0.598 $\pm$ 0.033            & 0.687 $\pm$ 0.019          & 0.758 $\pm$ 0.017           & 0.782 $\pm$ 0.018           & 0.811 $\pm$ 0.003           & 0.640 $\pm$ 0.007             & 0.704 $\pm$ 0.010          & 0.723 $\pm$ 0.008           & 0.777 $\pm$ 0.007           & 0.775 $\pm$ 0.004           \\ \hline
\multirow{3}{*}{Perceptron}                                & No Beamforming                                          & 0.657 $\pm$ 0.024            & 0.723 $\pm$ 0.026          & 0.733 $\pm$ 0.017           & 0.771 $\pm$ 0.036           & 0.691 $\pm$ 0.045           & 0.582 $\pm$ 0.058            & 0.710 $\pm$ 0.029          & 0.744 $\pm$ 0.037           & 0.754 $\pm$ 0.015           & 0.676 $\pm$ 0.073           \\ \cline{2-12} 
                                                           & DaS                                                 & 0.655 $\pm$ 0.036            & 0.735 $\pm$ 0.016          & 0.792 $\pm$ 0.025           & 0.801 $\pm$ 0.023           & 0.819 $\pm$ 0.032           & 0.659 $\pm$ 0.055            & 0.740 $\pm$ 0.018          & 0.763 $\pm$ 0.035           & 0.771 $\pm$ 0.046           & 0.769 $\pm$ 0.058           \\ \cline{2-12} 
                                                           & GSC                                                       & \textbf{0.670 $\pm$ 0.046}   & \textbf{0.746 $\pm$ 0.009} & 0.784 $\pm$ 0.019           & 0.807 $\pm$ 0.029           & 0.815 $\pm$ 0.025           & 0.649 $\pm$ 0.050            & 0.735 $\pm$ 0.025          & 0.769 $\pm$ 0.029           & 0.760 $\pm$ 0.058            & 0.767 $\pm$ 0.064           \\ \hline
\multirow{3}{*}{QDA}                                       & No Beamforming                                          & 0.380 $\pm$ 0.023            & 0.659 $\pm$ 0.030          & 0.601 $\pm$ 0.007           & 0.592 $\pm$ 0.023           & 0.554 $\pm$ 0.019           & 0.392 $\pm$ 0.003            & 0.660 $\pm$ 0.009          & 0.672 $\pm$ 0.009           & 0.615 $\pm$ 0.009           & 0.535 $\pm$ 0.011           \\ \cline{2-12} 
                                                           & DaS                                                 & 0.403 $\pm$ 0.038            & 0.659 $\pm$ 0.022          & 0.687 $\pm$ 0.016           & 0.678 $\pm$ 0.009           & 0.687 $\pm$ 0.052           & 0.428 $\pm$ 0.010            & 0.685 $\pm$ 0.006          & 0.803 $\pm$ 0.009           & 0.758 $\pm$ 0.004           & 0.754 $\pm$ 0.007           \\ \cline{2-12} 
                                                           & GSC                                                       & 0.400 $\pm$ 0.037              & 0.659 $\pm$ 0.025          & 0.680 $\pm$ 0.014            & 0.687 $\pm$ 0.028           & 0.697 $\pm$ 0.043           & 0.409 $\pm$ 0.010            & 0.678 $\pm$ 0.003          & 0.798 $\pm$ 0.007           & 0.763 $\pm$ 0.003           & 0.737 $\pm$ 0.009           \\ \hline
\multirow{3}{*}{SGD}                                       & No Beamforming                                          & 0.529 $\pm$ 0.017            & 0.607 $\pm$ 0.015          & 0.653 $\pm$ 0.021           & 0.723 $\pm$ 0.019           & 0.699 $\pm$ 0.030           & 0.714 $\pm$ 0.023            & 0.754 $\pm$ 0.016          & 0.796 $\pm$ 0.009           & 0.819 $\pm$ 0.008           & 0.767 $\pm$ 0.029           \\ \cline{2-12} 
                                                           & DaS                                                 & 0.565 $\pm$ 0.018            & 0.641 $\pm$ 0.017          & 0.731 $\pm$ 0.020           & 0.807 $\pm$ 0.015           & 0.822 $\pm$ 0.004           & 0.708 $\pm$ 0.012            & 0.784 $\pm$ 0.011          & 0.824 $\pm$ 0.015           & 0.853 $\pm$ 0.007           & 0.824 $\pm$ 0.012           \\ \cline{2-12} 
                                                           & GSC                                                       & 0.577 $\pm$ 0.017            & 0.634 $\pm$ 0.009          & 0.737 $\pm$ 0.024           & 0.817 $\pm$ 0.012           & 0.817 $\pm$ 0.007           & 0.720 $\pm$ 0.007             & 0.775 $\pm$ 0.011          & 0.830 $\pm$ 0.007           & 0.841 $\pm$ 0.009           & 0.824 $\pm$ 0.012           \\ \hline
\multirow{3}{*}{SVM}                                       & No Beamforming                                          & 0.548 $\pm$ 0.014            & 0.700 $\pm$ 0.024          & 0.773 $\pm$ 0.016           & 0.820 $\pm$ 0.011           & 0.811 $\pm$ 0.009           & 0.775 $\pm$ 0.014            & 0.775 $\pm$ 0.012          & 0.843 $\pm$ 0.004           & 0.840 $\pm$ 0.007           & 0.803 $\pm$ 0.004           \\ \cline{2-12} 
                                                           & DaS                                                 & 0.561 $\pm$ 0.022            & 0.733 $\pm$ 0.024          & \textbf{0.803 $\pm$ 0.007}  & 0.859 $\pm$ 0.018           & 0.851 $\pm$ 0.009           & \textbf{0.819 $\pm$ 0.013}   & \textbf{0.828 $\pm$ 0.008} & \textbf{0.847 $\pm$ 0.010}  & 0.855 $\pm$ 0.013           & 0.849 $\pm$ 0.007           \\ \cline{2-12} 
                                                           & GSC                                                       & 0.563 $\pm$ 0.011            & 0.740 $\pm$ 0.022          & 0.803 $\pm$ 0.019           & \textbf{0.859 $\pm$ 0.016}  & 0.847 $\pm$ 0.012           & 0.817 $\pm$ 0.012            & 0.824 $\pm$ 0.009          & 0.845 $\pm$ 0.011           & \textbf{0.860 $\pm$ 0.015}  & \textbf{0.853 $\pm$ 0.007}  \\ \hline
\multirow{3}{*}{Tree}                                      & No Beamforming                                          & 0.552 $\pm$ 0.074            & 0.613 $\pm$ 0.056          & 0.704 $\pm$ 0.019           & 0.729 $\pm$ 0.039           & 0.679 $\pm$ 0.041           & 0.651 $\pm$ 0.031            & 0.754 $\pm$ 0.026          & 0.798 $\pm$ 0.037           & 0.742 $\pm$ 0.045           & 0.668 $\pm$ 0.013           \\ \cline{2-12} 
                                                           & DaS                                                 & 0.601 $\pm$ 0.033            & 0.687 $\pm$ 0.041          & 0.786 $\pm$ 0.034           & 0.790 $\pm$ 0.033           & 0.811 $\pm$ 0.003           & 0.657 $\pm$ 0.028            & 0.775 $\pm$ 0.034          & 0.819 $\pm$ 0.044           & 0.822 $\pm$ 0.026           & 0.796 $\pm$ 0.029           \\ \cline{2-12} 
                                                           & GSC                                                       & 0.619 $\pm$ 0.036            & 0.685 $\pm$ 0.043          & 0.786 $\pm$ 0.036           & 0.796 $\pm$ 0.023           & 0.815 $\pm$ 0.018           & 0.662 $\pm$ 0.037            & 0.779 $\pm$ 0.044          & 0.819 $\pm$ 0.026           & 0.826 $\pm$ 0.016           & 0.798 $\pm$ 0.033           \\ \hline
\end{tabular}
        }

    \label{tab:AccSNR}
\end{table*}
    
    After simulating the microphones array with all dataset's audios and for several SNR values, three resulting signals were obtained:
    
    \begin{enumerate}
        \item \textbf{Without beamforming:} consisted of the sum of the signals from the four simulated microphones, without temporal displacement;
        
        \item \textbf{With DaS:} the delays between the microphones were repaired before the sum of the signals was performed.
        
        \item \textbf{With GSC:} the signal acquired after the DaS from the previous step was used to proceed with the GSC.
    \end{enumerate}

    
    Using a hyper-parameter configuration that provides the best accuracy for classifiers for both original and augmented datasets, the calculated signals \textit{without beamforming}, \textit {with DaS}, and \textit{with GSC} were used in the classification test step, where, again, bootstrap cross-validation was employed.
    
    \begin{figure}[!htb]
        \centering
        \includegraphics[keepaspectratio, width=13cm]{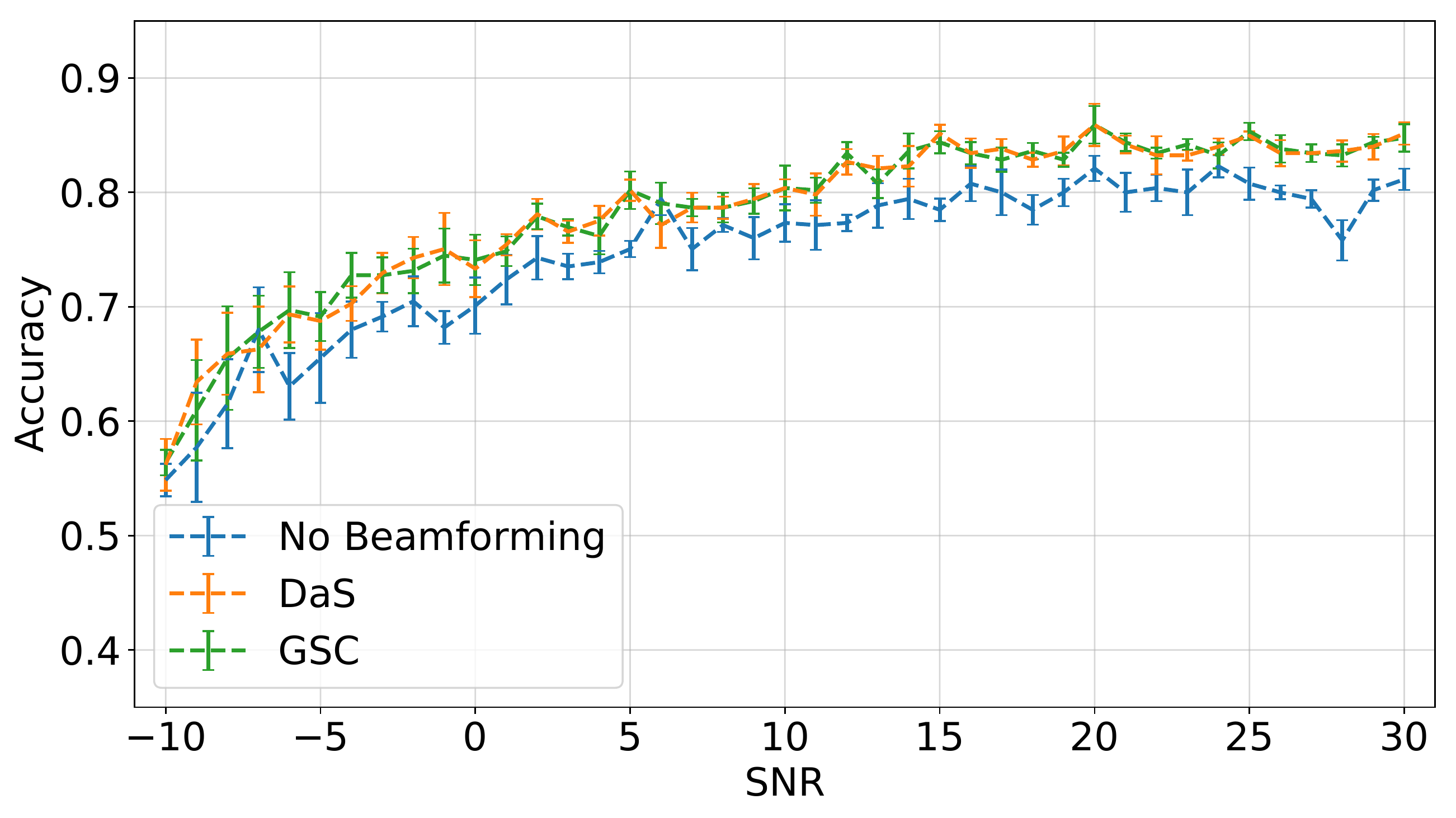}
        \caption{Mean accuracy and standard deviation of SVM as a function of SNR for different beamforming approaches when the classifier was trained with the original dataset.}
        \label{fig:grafico_ACCxSNR_original_dataset}
    \end{figure}
    
    \begin{figure}[!htb]
        \centering
        \includegraphics[keepaspectratio, width=13cm]{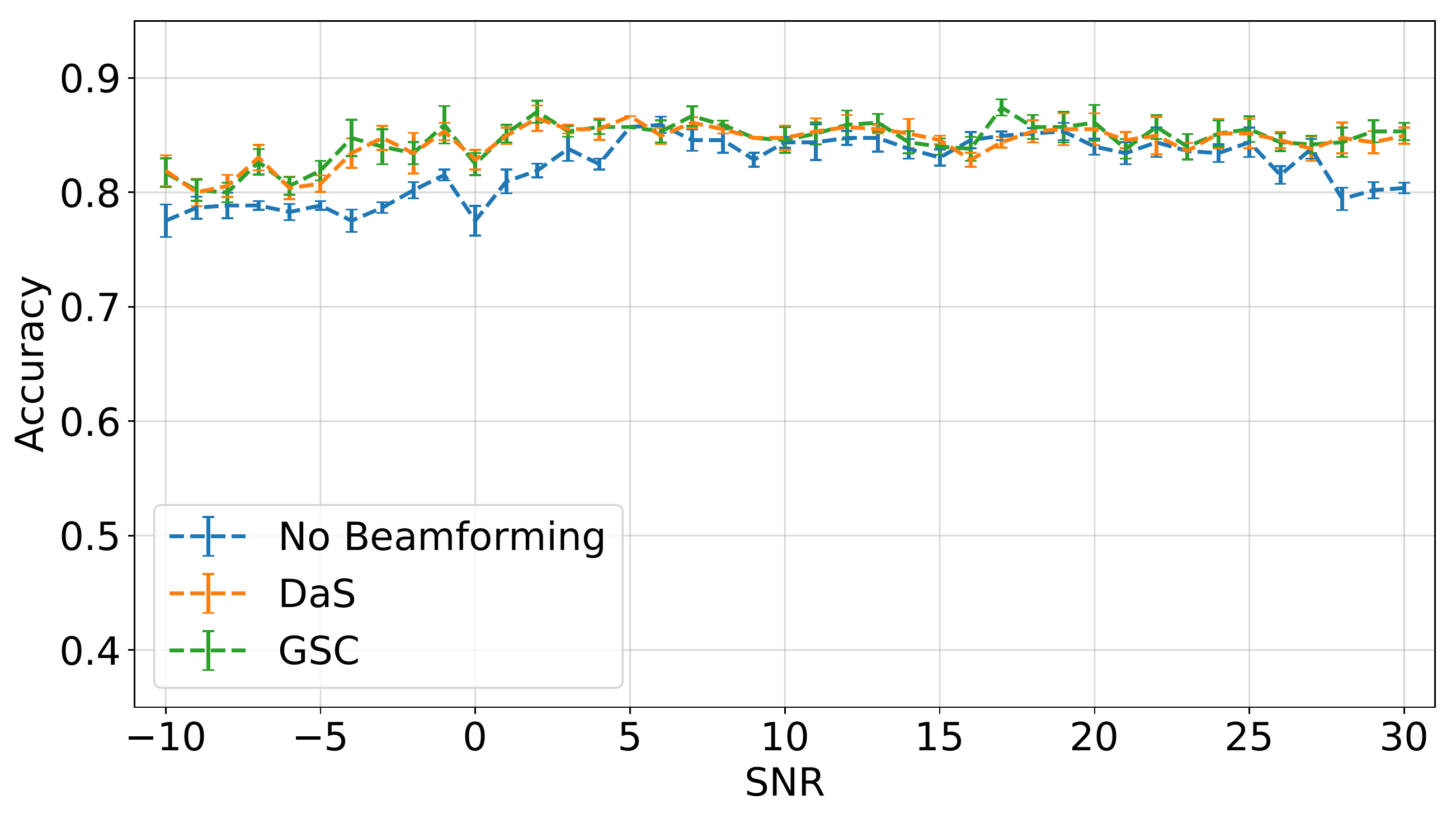}
        \caption{Mean accuracy and standard deviation of SVM as a function of SNR for different beamforming approaches when the classifier was trained with the augmented dataset.}
        \label{fig:grafico_ACCxSNR_data_augmentation}
    \end{figure}
    
    Table \ref{tab:AccSNR} brings comparative results between classifiers, beamforming and datasets for different SNR values. Figures \ref{fig:grafico_ACCxSNR_original_dataset} and \ref{fig:grafico_ACCxSNR_data_augmentation} show in more detail the same results for SVM, the classifier that obtained the best performance in accuracy.
    
    About the beamforming techniques applied, it was observed that they were successful in increasing the accuracy of all classifiers, in addition to having demonstrated consistency by presenting a practically constant improvement for all tested noise scenarios. The gain was approximately the same for both DaS and GSC.
    
    Regarding the data augmentation, there was an even more positive result: the classifiers showed a remarkably expressive increase in the noisiest scenarios and, in some cases, an irrelevant loss in the less noisy ones. Besides, the standard deviation has decreased, making the classifiers even more robust. When using SVM in combination with beamforming techniques, for example, the accuracy was greater than 80~\% in all scenarios. In general, the non-parametric classifiers, SVM and KNN, were the ones that managed to reach the highest accuracy values, corroborating the results obtained in \cite{16-SURAMPUDI2019}.

    
    \begin{figure}[!htb]
        \centering
        \includegraphics[keepaspectratio, width=10cm]{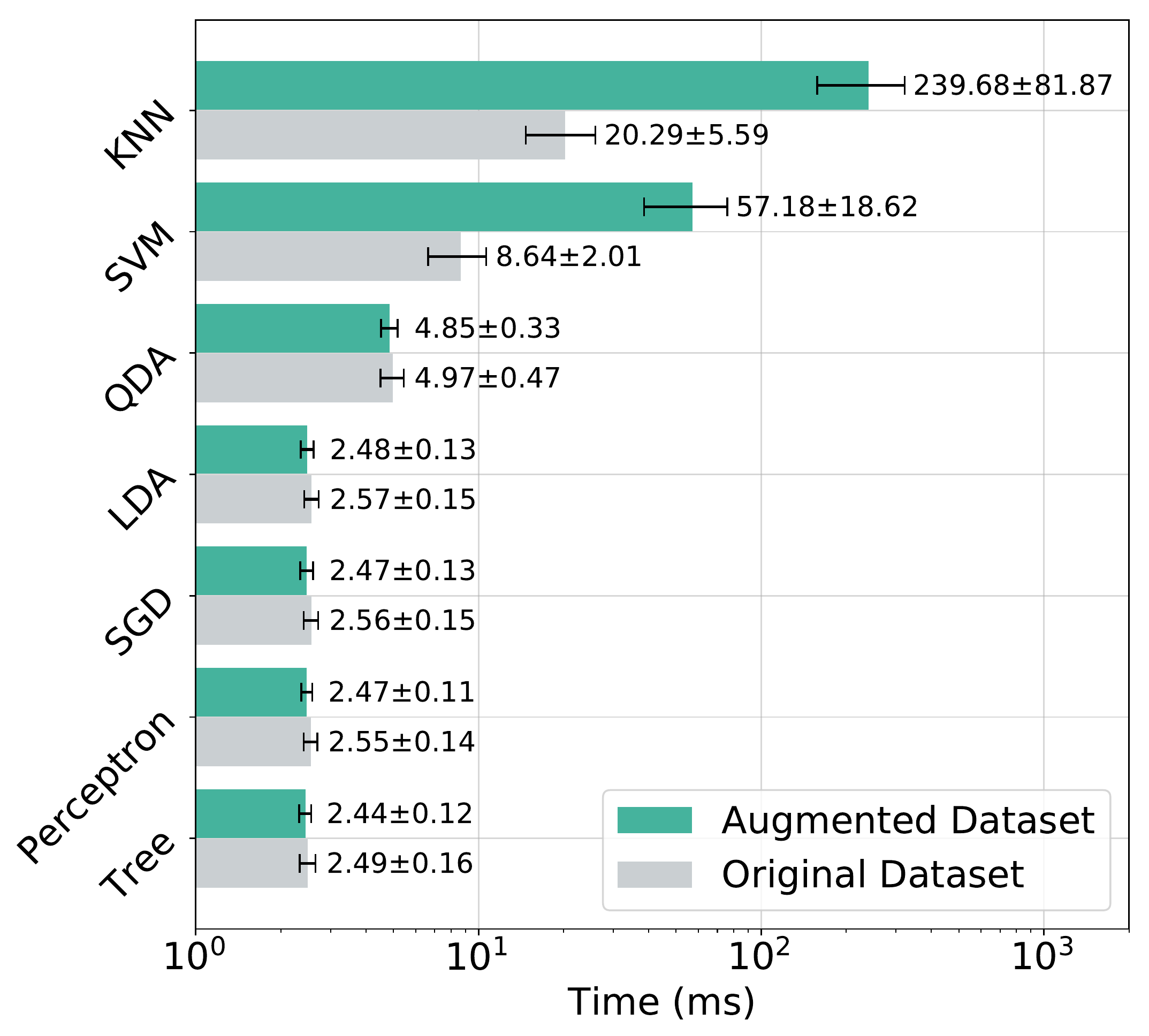}
        \caption{Average time and standard deviation required to classify all windows of a single audio when training was done with the original and augmented datasets.}
        \label{fig:tempoTodosClassificadores}
    \end{figure}
    
    
    As mentioned, the computational cost of each algorithm was also a concern. The tests indicated that beamforming techniques were not influenced as a result of the SNR, nor concerning the data augmentation since they were applied before classification. The average and standard deviation of processing time for these algorithms were:
    
    \begin{itemize}
        \item \textit{DaS}: (162.67 $\pm$ 99.99) ms.
        \item \textit{GSC}: (1279.69 $\pm$ 802.71) ms.
    \end{itemize}
    
    Figure \ref{fig:tempoTodosClassificadores} shows, in logarithmic scale, the mean and standard deviation of the time needed to classify all windows from the same audio, before and after data augmentation. As expected, parametric classifiers were not affected after this change. However, those that had the best accuracy score had sudden changes. The first, SVM, increased by 561.80~\%, while the second, KNN, 1081.27~\%. These changes may compromise future applications in real-time. 
    
    Analyzing the results, it appears that beamforming techniques were much more costly when compared to the classifiers, except for DaS, which was faster than KNN only after data augmentation. Specifically about GSC, since it obtained a processing time almost 10 times longer than DaS and did not present great accuracy differences, it's usage is not suggested in the next stages of this work.
    
    It is assumed that the most appropriate configuration in terms of accuracy is that which combines the augmented dataset, DaS, and SVM. Even so, concerning the processing time, SGD Classifier also deserves attention, since it has also achieved good accuracy results, and, unlike SVM, its processing time was not affected by data augmentation.

    \section{Conclusions}
    \label{sec:conclusao}
    
    The work presented a comparative study between seven classifiers and two beamforming techniques aiming to find an efficient configuration for the construction of an audio-based public security system. The results revealed that the best performance in accuracy was obtained with the combination of DaS and SVM algorithms. Furthermore, Data Augmentation guaranteed accuracy above 80~\% for all examined SNR values. Despite that, SGD Classifier conferred more satisfactory results in processing time, although the classification performance was slightly lower. Besides, it was found that the processing time of beamforming was much higher than the classifiers', which may compromise future real-time applications if they do not receive improvements.
    
    In future work, it is advisable to investigate other beamforming techniques and perform analyses with other databases, including noise bases. Also, the aim is to implement the system discussed on an embedded platform that allows the coupling of an array of microphones, and that can run in real-time.

\bibliographystyle{unsrt}  
\bibliography{references}  

\end{document}